\documentstyle[12pt,psfig]{article}
\makeatletter
\def\eqalign#1{\null\vcenter{\def\\{\cr}\openup\jot\m@th
  \ialign{\strut$\displaystyle{##}$\hfil&$\displaystyle{{}##}$\hfil
      \crcr#1\crcr}}\,}
\makeatother
\begin{document}
\bigskip
\bigskip
\bigskip
\begin{center}
{\Large\bf 
Asymptotic distribution of zeros of polynomials satisfying 
difference equations. 
}\\
\bigskip
\bigskip
\bigskip
\bigskip
{\large I. V. Krasovsky}\\
\bigskip
Max-Planck-Institut f\"ur Physik komplexer Systeme\\
N\"othnitzer Str. 38, D-01187, Dresden, Germany\\
E-mail: ivk@mpipks-dresden.mpg.de\\
\medskip
and\\
\medskip
B.I.Verkin Institute for Low Temperature Physics and Engineering\\
47 Lenina Ave., Kharkov 310164, Ukraine.
\end{center}
\bigskip
\bigskip
\bigskip

\noindent
{\bf Abstract.}
We propose a way to find the asymptotic distribution of zeros of orthogonal 
polynomials $p_n(x)$ satisfying a difference equation of the form
\[
B(x)p_n(x+\delta)-C(x,n)p_n(x)+D(x)p_n(x-\delta)=0.
\]
We calculate the asymptotic distribution of zeros and asymptotics of 
extreme zeros of the Meixner and Meixner-Pollaczek polynomials.
The distribution of zeros of Meixner polynomials shows some delicate features.
We indicate the relation of our approach to the WKB technique and to the
approach based on the Nevai-Ullman distribution.

\newpage
\section{Introduction}
The hypergeometric orthogonal polynomials $p_n(x)$ that appear in the Askey 
scheme [\ref{Koe}] satisfy either differential equations or difference
equations of the form:
\begin{equation}
B(x)p_n(x+\delta)-C(x,n)p_n(x)+D(x)p_n(x-\delta)=0,
\end{equation}
where $\delta$ is equal to $1$ or $i$ when the polynomials are orthogonal with
a discrete or continuous weight $\alpha(x)$, respectively.

As is well known (see, e.g., [\ref{Chi}]), the zeros of $p_n(x)$ are real and
simple. In the present paper, we will propose a procedure to calculate 
the asymptotic distribution of the zeros of $p_n(nz)$ as $n\to\infty$ by 
applying a form of Bethe ansatz to equation (1).\footnote{We will consider
infinite systems of orthogonal polynomials.} 
The difference between the cases $\delta=i$ and $\delta=1$ will be clearly
visible.

Note that although we focus on the orthogonal polynomials belonging to the
Askey scheme, our results have a more general application.
In fact, we need only the two conditions: 1) polynomials $p_n(x)$ satisfy
the difference equation (1) (or a certain type of $q$-difference equation
[\ref{sf-harp}]); 2) the density of zeros is 
piece-wise smooth on a one-dimensional manifold. For the polynomials
in the Askey scheme, the latter manifold is just the real line.
For another example, where the manifold is the unit circle, see
[\ref{sf-harp}].
    
In this work, we are interested in zeros of $p_n(x)$ only to the main order 
in $n$. However, the method we apply should remain useful for extraction 
of the further terms in the asymptotic series.

Other approaches to the calculation of the asymptotic distribution
of zeros of orthogonal polynomials are based on the knowledge of the 
recurrence relation, or the weight function $d\alpha(x)/dx$, or the asymptotic
behaviour of polynomials themselves (see [\ref{Lub}] for a review).

Our approach is closely related to the semiclassical (WKB) analysis of
differential (difference) equations. WKB techniques easily yield the more 
``rough'' features of the distribution of zeros such as their density.
Throughout the paper, we indicate how some of our results can be obtained by
the WKB method.

We will consider in detail the cases of the Meixner polynomials 
(orthogonal on the set $\{0,1,2,\dots\}$, $\delta=1$) and the Meixner-Pollaczek
polynomials (orthogonal on the set $(-\infty,\infty)$, $\delta=i$).
The asymptotics of zeros of the Meixner-Pollaczek polynomials has 
been investigated in [\ref{VACA},\ref{Ism}] by other methods. 
Therefore, the reader can compare several approaches.

The zeros of $p_n(x)$ are intimately related to the points of increase of the
weight (see, e.g., [\ref{Chi}]). If the moment problem corresponding
to the polynomials $p_n(x)$ is determinate, the points of increase of the
weight coincide with the closure of the set of zeros of $p_n(x)$,
$n\to\infty$. For the Meixner and the Meixner-Pollaczek polynomials the
latter fact implies that the density $\rho(z)$ of the zeros of $p_n(nz)$
at the point $z=0$ is $1$ and $\infty$, respectively. 
Naturally, the expressions for $\rho(z)$ we obtain have this property.
It is interesting that, moreover, for the Meixner polynomials $\rho(z)=1$ in
the {\it finite} vicinity of $z=0$. In this vicinity the distribution of zeros
is characterized by another function (simply related to $\rho(z)$) that 
gives the rate with which the
distance between zeros approaches the value 1 as $z\to 0$. This rate is
exponential in $n$. 

There is another interpretation of our results on the Meixner and
Meixner-Pollaczek polynomials. Consider a
symmetric tridiagonal (Jacobi) matrix with real matrix elements 
($J_{ik}=J_{ki}$,
$i,k=0,1,2,\dots$, $J_{ik}=0$ if $|i-k|>1$) such that the limits 
$\lim_{n\to\infty}J_{nn}/\varphi(n)=a$, 
$\lim_{n\to\infty}J_{n\;n+1}/\varphi(n)=b/2$ are finite and not 
simultaneously zero and $\varphi(n)$ is a nondecreasing function.
All such matrices for $\varphi(n)=n$ are separated in two classes. 
The first class is comprised of matrices related to the Meixner
polynomials and the second, to Meixner-Pollaczek. The asymptotics of zeros of
polynomials is the asymptotics of eigenvalues of the corresponding truncated 
Jacobi matrices. Thus in this paper we have calculated, in particular,
the asymptotic distribution 
of eigenvalues of the above described Jacobi matrices for $\varphi(n)=n$.
On the other hand, the density of zeros in the case of $a=0$ is known for
any $\varphi(n)$ (with some limitations on the form of $\varphi(n)$). 
It is called Nevai-Ullman distribution [\ref{Nev},\ref{Ull}].
It can be easily generalized to the case of any real $a$, $b$. 
In particular for $\varphi(n)=n$, this gives an alternative derivation
of the density of zeros of the Meixner and Meixner-Pollaczek polynomials.  
However, this does not reproduce the delicate features of the zero
distribution of the Meixner polynomials in the vicinity of $z=0$.

The plan of this work is the following. In Section 1 we introduce the Bethe
ansatz technique in the simplest case of polynomials satisfying differential
equations. We obtain the asymptotic density of zeros and compare the results
with those of a more straightforward approach. 

In Section 2 we consider the difference equation (1) with $\delta=i$.
It is satisfied by polynomials orthogonal with a continuous weight:
Meixner-Pollaczek, continuous Hahn, continuous dual Hahn, Wilson.
After getting the general solution, we focus on Meixner-Pollaczek as the other
polynomials have, to the main order in $n$, 
the density of zeros simpler than that of Meixner-Pollaczek.

In Section 3 we consider the more interesting case of equation (1) with 
$\delta=1$. Such equation is satisfied by the Meixner and Charlier polynomials.
We will consider only them, as other polynomials with discrete weights
in the Askey scheme are finite systems.

In Section 4 we consider the asymptotic eigenvalue
problem for symmetric tridiagonal matrices.

\section{Differential equation}
We shall start with the case of a differential equation as it is the simplest
one, and it already shows some features of the method.

Consider a system of polynomials satisfying the following differential
equation:\footnote{Here we restrict our attention to second-order
differential equations, but it is easy to generalize our approach to
higher-order equations.}
\begin{equation}
a(x)y^{''}_n+b(x)y^{'}_n+c(x,n)y_n=0.\label{de}
\end{equation}
Here $y_n(x)$ is a polynomial of exactly degree $n$, functions $a(x)$,
$b(x)$, $c(x,n)$, where only the last one depends on $n$, are
smooth with possible singular points that do not coincide with any zeros of
$p_n(x)$.  We shall assume that there
is such a constant $\mu>0$ that $c(x,n)\sim c_\infty(x)n^{2\mu}$ as
$n\to\infty$. Such equations are satisfied, e.g., by the Jacobi, Laguerre, 
Hermite polynomials. These polynomials are orthogonal with an absolutely 
continuous weight
on an interval $\cal A$ (unbounded in the case of Laguerre and
Hermite) and the zeros of $y_n(x)$ become everywhere dense in 
$\cal A$ as $n\to\infty$.

Let  $x_1<x_2<\cdots<x_n$ be the zeros of $y_n(x)$. Up to a constant factor,
we can write $y_n(x)$ in the form $y_n(x)=\prod_{k=1}^n (x-x_k)$.
Substituting this expression in (\ref{de})
and dividing the equation by $y_n(x)$, we get
\begin{equation}
a(x)\sum_{k=1}^n\sum_{i=1\atop i\neq k}^n\frac1{(x-x_k)(x-x_i)}+
b(x)\sum_{k=1}^n\frac1{x-x_k}+c(x,n)=0.\label{4}
\end{equation}
Obviously, the singularities of the function at the left-hand side must 
vanish.
Equating the residues in the simple poles $x_k$, $k=1,\dots,n$, to zero,
we obtain the following system:
\begin{equation}
2a(x_k)\sum_{i=1\atop i\neq k}^n\frac1{x_k-x_i}+b(x_k)=0,\qquad
k=1,\dots,n.\label{d0}
\end{equation}
Equations (\ref{d0}) can also be obtained
by dividing (\ref{de}) by $y^{'}_n(x)$ and putting $x=x_k$. The quantities 
$\sigma_{pk}=\sum_{i=1\atop i\neq k}^n 1/(x_k-x_i)^p$, 
$p=1,2,\dots$ and related
ones were calculated in [\ref{ABC},\ref{Case},\ref{Ahm}]. Analogous
algebraic relations for solutions of some $q$-difference equations are given
in [\ref{WZ}]. Because of the analogy to the theory of integrable models,
the equations of type (\ref{d0}) are called Bethe ansatz equations.
For our purposes, it will be sufficient to know only the
asymptotics of $\sigma_{1k}$ and $\sigma_{2k}$ as $n\to\infty$.  

Let us now differentiate (\ref{de}) with respect to $x$ and apply the
same polynomial-ansatz technique to the resulting equation.
Then instead of (\ref{d0}) we get
\begin{eqnarray}
\eqalign{
3a(x_k)\left\{\sum_{i=1\atop i\neq k}^n\frac1{x_k-x_i}
\sum_{j=1\atop j\neq k}^n\frac1{x_k-x_j}-
\sum_{i=1\atop i\neq k}^n\frac1{(x_k-x_i)^2}\right\}+\\
2(a'(x_k)+b(x_k))\sum_{i=1\atop i\neq k}^n\frac1{x_k-x_i}+
c(x_k,n)+b'(x_k)=0,\qquad k=1,\dots,n.}\label{d1}
\end{eqnarray}
Now divide (\ref{d1}) by $n^{2\mu}$ and take the limit as $n\to\infty$ on
condition that the point $x=\lim_{n\to\infty}x_{k(n)}$, $x\in\cal A$,
is fixed. Then,
using the fact that, as follows from (\ref{d0}),
\begin{equation}
\lim_{n\to\infty}\frac1{n^\sigma}\sum_{i=1\atop i\neq k}^n
\frac1{x_k-x_i}=0\label{inf}
\end{equation}
for any $\sigma>0$, we obtain from (\ref{d1})
\begin{equation}
\lim_{n\to\infty}\frac1{n^{2\mu}}\sum_{i=1\atop i\neq k}^n
\frac1{(x_k-x_i)^2}=\frac{c_\infty(x)}{3a(x)}.\label{n2}
\end{equation}

Let us assume that the zeros are asymptotically spaced as $1/n^\mu$ and their
asymptotic density is a piece-wise smooth function. 
Hence for $|i-k|\le M$ we have 
\begin{equation}
x_i-x_k=(i-k)\gamma_k/n^\mu\label{ik}
\end{equation}
up to at most 
$O((M/n^\mu)^2)$ as $M,n\to\infty$ in such a way that $M/n^\mu\to 0$.
For brevity, we shall denote the limit $M,n\to\infty$, $M/n^\mu\to 0$
just by $\lim$.
In (\ref{ik})  $\lim\gamma_{k(n)}=\gamma(x) > 0$.   
The asymptotic density of zeros is, obviously, 
$\tilde{\rho}(x)=\lim 1/\gamma_{k(n)}$. 
Using (\ref{ik}) we rewrite the l.h.s. of (\ref{n2}) in the form:
\begin{equation}
\lim\frac1{n^{2\mu}}\sum_{i=1\atop i\neq k}^n
\frac1{(x_k-x_i)^2}=2\tilde{\rho}(x)^2\sum_{i=1}^\infty\frac1{i^2}=
\frac{\pi^2}3\tilde{\rho}(x)^2,\label{9}
\end{equation}
So putting (\ref{n2}) and (\ref{9}) together, we have
\begin{equation}
\tilde{\rho}(x)=\frac1{\pi}\sqrt{\frac
{c_\infty(x)}{a(x)}}\label{rho0}
\end{equation}
on the set ${\cal A}$.
It is not necessary to justify (\ref{rho0}) further as 
this result is, of course, well known. It is related to 
the semiclassical approximation in the theory of Schr\"odinger
equation (see, e.g., [\ref{LL3}]). In a few words, the
usual way to obtain (\ref{rho0}) from (\ref{de}) is the following.
Consider a small
$\sigma$-neighbourhood of a point $x$. If $a(x)c_\infty(x)>0$ then,
starting with a sufficiently large number $n$, the characteristic values
of the equation (\ref{de}) (considered as an equation with constant 
coefficients:  $a(x)$, $b(x)$, $c(x,n)$ being sufficiently smooth),
 $\omega_\pm=\{-b(x)\pm\sqrt{b(x)^2-4a(x)c(x,n)}\}/(2a(x))$, have the
imaginary part growing as $n^{\mu}$. Hence, it is easy to conclude
that the asymptotic distance between consecutive zeros is
$x_{k+1}-x_k\sim\gamma_k/n^{\mu}$,
$\gamma_k=\pi\sqrt{\frac{a(x)}{c_\infty(x)}}+o(1)$.

For the polynomials orthogonal on an unbounded interval, another type of the
density of zeros can be introduced. Take, for example, the Hermite
polynomials. Their zeros are symmetric with respect to the origin, and
the largest (smallest) zero $x_n\sim\sqrt{2n}$ ($x_1\sim-\sqrt{2n}$).
Equation (\ref{d0}) in this case has the form:
\begin{equation}
\sum_{i=1\atop i\neq k}^n\frac1{x_k-x_i}=x_k,\qquad
k=1,\dots,n.\label{H}
\end{equation}
Changing the variable $x_i=z_i\sqrt{2n}$ and then taking the limit
$n\to\infty$, we get [\ref{CP}] 
\begin{equation}
{1\over 2}{\rm V.p.}\int_{-1}^1\frac{\rho(\omega)d\omega}{z-\omega}=z,
\qquad z\in(-1,1),
\label{IH}
\end{equation}
where the density of zeros $\rho(\omega)=\lim_{n\to\infty} 1/(z_{i+1}-z_i)n$,
$\omega=\lim_{n\to\infty}z_{i(n)}$.
Solving (\ref{IH}) and using the normalization condition
$\int_{-1}^1 \rho(\omega)d\omega=1$, we get
\begin{equation}
\rho(z)=\frac2\pi\sqrt{1-z^2}.\label{r}
\end{equation}
This way to derive (\ref{r}) was proposed in [\ref{CP}]. Of course, there
are other ways to find  $\rho(z)$.
Expression (\ref{r}) is valid for a contracted to $[-1,1]$ density of zeros of 
a large class of orthogonal polynomials (to which Hermite polynomials belong).
It is a particular case of the Nevai-Ullman
distribution (see Section 5) for $\varphi(n)=\sqrt{n}$.

Note that (\ref{rho0}) for Hermite polynomials gives
$\tilde{\rho}(x)=\sqrt{2}/\pi$. 

In the next Section we shall see how the technique we used to obtain
(\ref{rho0}) generalizes to the difference equations.

\section{Difference equation. Continuous weight.}
Consider polynomials in the Askey scheme orhtogonal with a continuous weight
on an unbounded interval and satisfying (1). In this case $\delta=i$.
Take the n'th polynomial and make the change of variable $x=zn$. Then
$p_n(x)=\prod_{k=1}^n(x-x_k)=n^ny(z)$, where
$y(z)=\prod_{k=1}^n(z-z_k)$. As before, we assume  $x_1<x_2<\cdots<x_n$.
If $C(x,n)\sim c(z)n^\mu$ as $n\to\infty$,
then set $b(z)=\lim_{n\to\infty}B(zn)/n^\mu$,
$d(z)=\lim_{n\to\infty}D(zn)/n^\mu$.
Thus, to the main order in $n$, (1) can be written in the form:
\begin{equation}
b(z)y(z+i/n)-c(z)y(z)+d(z)y(z-i/n)=0.\label{di}
\end{equation}
Substituting here $y(z)=\prod_{k=1}^n(z-z_k)$ and evaluating the equation
at the zeros $z_m$, $m=1,2,\dots,n$, we get:
\begin{equation}
\prod_{k=1}^n\frac{z_m-z_k+i/n}{z_m-z_k-i/n}=-\frac{d(z_m)}{b(z_m)},
\qquad m=1,2,\dots,n.
\label{ba11}
\end{equation}

Now assume that the distance between zeros $x_k$ is of order 1 (which is
natural to expect from (1)), and the asymptotic density 
$\rho(z_m)=1/(x_m-x_{m-1})$ of zeros 
becomes a piece-wise smooth function of $z=\lim_{n\to\infty}z_{m(n)}$  
as $n\to\infty$. Hence for $|p|<M$
\begin{equation}
z_m-z_{m-p}=\frac{p}{\rho(z)n}\label{nzeros}
\end{equation}
up to at most $O(M^2/n^2)$ as $M,n,n/M\to\infty$. As before, we denote this
limit by ``lim''. Applying ``lim'' to (\ref{ba11}) we have
\begin{equation}
\lim\left(-\prod_{p=1}^M\frac{{p\over\rho(z)}+i}{{p\over\rho(z)}-i}\,
\frac{{-p\over\rho(z)}+i}{{-p\over\rho(z)}-i}
\prod_k\mathop{^{'}}\frac{1+\frac{i/n}{z_m-z_k}}
{1-\frac{i/n}{z_m-z_k}}\right)=
-\frac{d(z)}{b(z)},
\end{equation}
where the prime indicates that the product is taken over $k$ outside the range
$m-M,\dots,m+M$.
The product over $p$ gives 1 in the limit. Since $i/n(z_m-z_k)$ in the second
product is small (of order $1/M$ or less), we can perform the small-parameter
expansion in it. Replacing the second product by the exponent of its logarithm
we get $\lim\exp[(2i/n)\sum_k\mathop{^{'}}1/(z_m-z_k)]=d(z)/b(z)$ which, in
turn, gives
\begin{equation}
\exp\left(-2i\, {\rm V.p.}\int_{-\infty}^{\infty}\frac{\rho(\omega)d\omega}
{\omega-z}\right)=\frac{d(z)}{b(z)}.
\label{ba11p}
\end{equation}

Note that $\rho(z)$ also satisfies the normalization condition 
\begin{equation}
\int_{-\infty}^{\infty}\rho(z)dz=1.\label{N}
\end{equation}

Instead of trying to solve (\ref{ba11p}) directly, we will do the following.
Replace $z$ by $z+i/n$ in (\ref{di}) and consider this new equation 
(cf. Section 2). Then we obtain instead of (\ref{ba11}):

\begin{equation}
\prod_{k=1}^n\frac{z_m-z_k+2i/n}{z_m-z_k+i/n}=\frac{c(z_m)}{b(z_m)},
\qquad m=1,2,\dots,n.
\label{ba12}
\end{equation}
Proceeding as before, we get for the l.h.s. of (\ref{ba12}):
\begin{equation}
\eqalign{
\lim\prod_{k=1}^n\frac{z_m-z_k+2i/n}{z_m-z_k+i/n}=
2\lim\prod_{s=0}^{[M/2]}\left\{1+\left(\frac{2\rho(z)}{2s+1}\right)^2\right\}
\exp\left({i\over n}\sum_k\mathop{^{'}}{1\over z_m-z_k}\right)=\\
2\cosh(\pi\rho(z)) \exp\left(-i\, {\rm V.p.}\int_{-\infty}^{\infty}
\frac{\rho(\omega)d\omega}{\omega-z}\right).}
\end{equation}
Thus, in the limit equation (\ref{ba12}) becomes
\begin{equation}
2\cosh(\pi\rho(z)) \exp\left(-i \,{\rm V.p.}\int_{-\infty}^{\infty}
\frac{\rho(\omega)d\omega}{\omega-z}\right)=\frac{c(z)}{b(z)}.
\label{ba12p}
\end{equation}

Putting (\ref{ba11p}) and (\ref{ba12p}) together, we obtain
\begin{equation}
\rho(z)=\frac 1\pi {\rm arccosh}\left|\frac{c(z)}{2\sqrt{b(z)d(z)}}\right|.
\label{rho}
\end{equation}
Thus $\rho(z)$ can be either (\ref{rho}) or zero. If we suppose that 
$\rho(z)$ is defined by (\ref{rho}) for all $z$ where 
$|c(z)/(2\sqrt{b(z)d(z)})|>1$ (hence $\rho(z)=0$ otherwise) then the
asymptotic values of the
largest and smallest zeros are obtained from the equations 
$c(z)/(2\sqrt{b(z)d(z)})=\pm 1$.

In the concrete cases we can verify if thus obtained $\rho(z)$ is indeed the
asymptotic density of zeros by substituting it in (\ref{ba11p}), (\ref{N}),
(\ref{ba12p}) and integrating. These concrete cases include
the Wilson, continuous Hahn, continuous dual Hahn, and Meixner-Pollaczek 
polynomials [\ref{Koe}].

The most interesting example is provided by the Meixner-Pollaczek polynomials.
They are orthogonal on the line $(-\infty,\infty)$  with the weight
function $d\alpha(x)/dx=\exp[(2\phi-\pi)x]|\Gamma(\lambda+ix)|^2$,
where $\lambda>0$, $0<\phi<\pi$. (Note that $d\alpha(x)$ can be
obtained [\ref{NUS}] from the difference equation). The coefficients in
equation (1) for these polynomials are the following:
$B(x)=e^{i\phi}(\lambda-ix)$, $C(x,n)=2i[(n+\lambda)\sin\phi-x\cos\phi]$,
$D(x)=-e^{-i\phi}(\lambda+ix)$, and $\delta=i$. 

Using the just obtained general results, we can immediately formulate a
theorem.

\noindent{\bf Theorem 1.} {\it
For the  Meixner-Pollaczek polynomials $p_n(zn)$ the asymptotic density of 
zeros is
\[
\rho(z)=\frac 1\pi {\rm arccosh}\left|\frac{\sin\phi}z-\cos\phi\right|,
\qquad {\rm if}\quad z\in[-\cot{\phi\over 2},\tan{\phi\over 2}],
\]
and $\rho(z)=0$ otherwise.} 
\bigskip

We note that the values $n\tan{\phi\over 2}$ and $-n\cot{\phi\over 2}$
are the asymptotics of the largest and smallest zeros of 
the  Meixner-Pollaczek polynomials $p_n(x)$, respectively.
There can be no isolated zeros $z_m$ outside the interval 
$[-\cot{\phi\over2},\tan{\phi\over 2}]$ as the Bethe ansatz equations
(\ref{ba11},\ref{ba12}) would not hold for such zeros. We can also use here
another argument. Consider the Gerschgorin circles of the Jacobi matrix
associated with $p_n(x)$ (see Section 5). We then find that all zeros must be 
inside the interval $[-\cot{\phi\over2},\tan{\phi\over 2}]$.

On the assumption of smoothness of $\rho(z)$, it is easy to prove the 
theorem by verifying
(\ref{ba11p}), (\ref{N}), and (\ref{ba12p}). In particular, we have
\begin{equation}
{\rm V.p.}\int_{-\cot{\phi\over 2}}^{\tan{\phi\over 2}}
\frac{\rho(\omega)d\omega}{\omega-z}=
\cases{\phi-\pi, & $0<z<\tan{\phi\over 2}$\cr
\phi, & $-\cot{\phi\over 2}<z<0$}.
\label{int1}
\end{equation}

As such, our results for the Meixner-Pollaczek polynomials are not new. 
In the implicit form they are contained in [\ref{VACA},\ref{Ism}].

In the next section we will use Bethe-ansatz techniques to 
indicate some interesting features of the distribution of zeros of Meixner
polynomials in the region where the asymptotic density of zeros is constant.
However, the formulas for the density itself ((\ref{rho}) and (\ref{rho2}))
are easier to obtain using the WKB approach which we now briefly describe.  
Suppose that $y(z)$ has the asymptotic form $y(z)\sim \exp(nS_0(z)+
S_1(z)+S_2(z)/n+\cdots)$, where $S_j(z)$, $j=0,1,\dots$ are continuous,
and substitute this expression into (\ref{di}). Decompose $S_j(z\pm i/n)$
into Taylor series and collect the terms of the same order in $n$ in the
equation. Solution of the equation to the main order in $n$ gives $S_0(z)$
which, in turn, yields the density of zeros. We see that this approach gives
not only the density of zeros but also the asymptotic expression for $y(z)$.

\section{Difference equation. Discrete weight.}
The Meixner and Charlier polynomials orhtogonal on the set $\{0,1,2,\dots\}$
satisfy equation (1) with $\delta=1$. Using the same notation as in Section 3,
we reduce (1) to the form:
\begin{equation}
b(z)y(z+1/n)-c(z)y(z)+d(z)y(z-1/n)=0.\label{did}
\end{equation}
Evaluating it
at the zeros $z_m$, $m=1,2,\dots,n$, we get:
\begin{equation}
\prod_{k=1}^n\frac{z_m-z_k+1/n}{z_m-z_k-1/n}=-\frac{d(z_m)}{b(z_m)},
\qquad m=1,2,\dots,n.
\label{ba21}
\end{equation}

For the distance between nearest zeros we still have the expression
(\ref{nzeros}). However, as we can notice from (\ref{ba21}), there appears a
new important feature. We mentioned in the introduction that in the present
case $\rho(z)\to 1$ as $z\to 0$. Hence we can expect that
$z_m-z_{m-1}-1/n=o(1/n)$ and  $z_m-z_{m+1}+1/n=o(1/n)$ for $z_m$ close to zero.
Therefore, to the main order in $n$, the ratio of these quantities is an
unknown if $\rho(z_m)=1$ and equal to $-1$ if $\rho(z_m)\ne
1$.\footnote{Strictly speaking, we also had to consider the cases when 
$\rho(z_m)=s$, where $s$ is an integer greater that 1. In this case
$z_m-z_{m\pm s}\pm 1/n=o(1/n)$. However, we shall see that $\rho(z)\le 1$ for
all $z$ {\it at least} for the Meixner and Charlier polynomials.}
We denote $\Delta_{m-1}= z_m-z_{m-1}-1/n$. 
Now we can proceed as in the previous section. 
\begin{equation}
\eqalign{
\lim\prod_{k=1}^n\frac{z_m-z_k+1/n}{z_m-z_k-1/n}=
\lim\left(-\frac{z_m-z_{m+1}+1/n}{z_m-z_{m+1}-1/n}\,
\frac{z_m-z_{m-1}+1/n}{z_m-z_{m-1}-1/n}\times\right.\\
\left.\prod_{p=2}^M\frac{{p\over\rho(z)}+1}{{p\over\rho(z)}-1}\,
\frac{{-p\over\rho(z)}+1}{{-p\over\rho(z)}-1}
\prod_k\mathop{^{'}}\frac{1+\frac{1/n}{z_m-z_k}}
{1-\frac{1/n}{z_m-z_k}}\right)=\\
-\lim\left(\frac{\Delta_m}{\Delta_{m-1}}
\exp\left[{2\over n}\sum_k\mathop{^{'}}{1\over z_m-z_k}\right]\right)
=-\frac{d(z)}{b(z)}.}
\end{equation}

Thus, in the limit (\ref{ba21}) becomes
\begin{equation}
\chi(z)\exp\left(-2\, {\rm V.p.}\int_{-\infty}^\infty\frac{\rho(\omega)d\omega}
{\omega-z}\right)=\frac{d(z)}{b(z)},
\label{ba21p}
\end{equation}
where $\chi(z)=\lim(\Delta_m/\Delta_{m-1})$, $z=\lim_{n\to\infty}z_{m(n)}$.

Remember that $\rho(z)$ satisfies the normalization condition 
\begin{equation}
\int_{-\infty}^{\infty}\rho(z)dz=1.\label{N2}
\end{equation}

As in Section 3, change the variable  $z\to z+1/n$ in (\ref{did}). Evaluating
thus obtained equation at the zeros $z_m$, we have:
\begin{equation}
\prod_{k=1}^n\frac{z_m-z_k+2/n}{z_m-z_k+1/n}=\frac{c(z_m)}{b(z_m)},
\qquad m=1,2,\dots,n.
\label{ba22}
\end{equation}
In the limit (\ref{ba22}) becomes
\begin{equation}
\eqalign{
\lim\left\{\left(1+\frac{\Delta_{m+1}}{\Delta_m}\right)
\prod_{s=0}^{[M/2]}\left[1-\left(\frac{2\rho(z)}{2s+1}\right)^2\right]
\exp\left({1\over n}\sum_k\mathop{^{'}}{1\over z_m-z_k}\right)\right\}=
\frac{c(z)}{b(z)}}\label{prom}
\end{equation}
Since $\lim(\Delta_{m+1}/\Delta_m)=\chi(z)$, we may 
rewrite (\ref{prom}) in the form:  
\begin{equation}
(1+\chi(z))
\cos(\pi\rho(z)) \exp\left(-{\rm V.p.}\int_{-\infty}^{\infty}
\frac{\rho(\omega)d\omega}{\omega-z}\right)=
\frac{c(z)}{b(z)}.\label{ba22p}
\end{equation}

From (\ref{ba21p}) and (\ref{ba22p}) we obtain a simple relation between 
$\rho(z)$ and $\chi(z)$:
\begin{equation}
(1+\chi(z))\cos(\pi\rho(z))={c(z)\over b(z)}\sqrt{\chi(z)b(z)\over d(z)}
.\label{rhochi}
\end{equation}
We know that $ \chi(z)=1$ whenever $\rho(z)$ is nonconstant.
Hence, if $\rho(z)$ is nonconstant, it satisfies the equation:
\begin{equation}
\cos(\pi\rho(z))=\frac{c(z)}{2b(z)}\sqrt{\frac{b(z)}{d(z)}}.\label{rho2}
\end{equation}
Obviously, this equation can hold only in the region where the absolute value
of the r.h.s. is no more than 1.
Having guessed $\rho(z)$ and $\chi(z)$ from (\ref{rhochi}), we can check 
if we are right by verifying (\ref{ba21p}), (\ref{ba22p}), and (\ref{N2}).
Thus we get the following theorem for the Meixner polynomials $p_n(x)$
for which [\ref{Koe}] $B(x)=(x+\beta)c$, $C(x)=x+(x+\beta)c+(c-1)n$, $D(x)=x$,
$\beta>0$, $0<c<1$.

\noindent{\bf Theorem 2.} {\it
For the  Meixner polynomials $p_n(zn)$ the asymptotic density of 
zeros is
\[
\rho(z)=\cases{1, & $z\in[0,\alpha^{-1}]$\cr
\frac 1\pi {\rm arccos}f(z), 
& $z\in[\alpha^{-1},\alpha]$\cr
0, & $z\notin[0,\alpha]$},
\]
where $0\le\arccos z\le\pi$, and
\[
\alpha=\frac{1+\sqrt{c}}{1-\sqrt{c}}, \qquad 
f(z)=\frac{(c+1)z+c-1}{2\sqrt{c}z}.
\]

The function

\[
\ln\chi(z)=\cases{
2\;{\rm arcosh}\;|f(z)|, & $z\in[0,\alpha^{-1}]$\cr
0, & $z\in[\alpha^{-1},\alpha]$}.
\]}
\bigskip

{\it Remarks.}
Since in the limit $\Delta_{m-1}=\exp[-(k/n)n\ln\chi(z)]\Delta_{m-1+k}$, and
$\chi(z)>1$ for $z\in[0,\alpha^{-1}]$,
the function $\ln\chi(z)$ describes the exponential rate with which zeros $x_k$
approach integer values as $z$ decreases from the point $\alpha^{-1}$.

The function $\rho(z)$ is continuous on the set $(0,\infty)$.
Its derivative, however, is discontinuous at the points $\alpha^{-1}$ and
$\alpha$.

The value $n\frac{1+\sqrt{c}}{1-\sqrt{c}}$ is the asymptotics of
the largest zero $x_{\rm max}$ of the Meixner polynomials $p_n(x)$. 
Note that the 
inequality $x_{\rm max}<n\frac{1+\sqrt{c}}{1-\sqrt{c}}+o(n)$ is required 
by the position of the Gerschgorin circles of the associated Jacobi matrix.
For further and more precise inequalities for zeros of $p_n(x)$ see
[\ref{IM}]. 

By the substitution $\phi=\ln\sqrt{c}$, we can make similarities with the
Theorem 1 more transparent. Indeed, $f(z)=(\sinh\phi)/z+\cosh\phi$, 
$\alpha=-\coth(\phi/2)$. One might have expected such similarities, because
the Meixner-Pollaczek polynomials are obtained from the Meixner polynomials by
continuing the latter in the variable $x$ and parameter $c$ into the complex
plane [\ref{NUS}]. 

The asymptotic equation (\ref{did}) for the Charlier polynomials can be 
regarded as a particular case of the one for the Meixner 
polynomials when $c\to 0$.

In the next section we will consider another approach (initiated in
[\ref{Nev}]) to calculation of the
asymptotic density of zeros of orthogonal polynomials. It is based on the
study of the recurrence relation rather that the difference equation.
This approach allows us to give an alternative derivation of 
$\rho(z)$ for Meixner and Meixner-Pollaczek polynomials (Theorem 1
and the first part of Theorem 2).

\section{Eigenvalue density of Jacobi matrices}
Consider a
symmetric Jacobi matrix $J$ with real matrix elements 
($J_{ik}=J_{ki}$, $i,k=0,1,2,\dots$, $J_{ik}=0$ if $|i-k|>1$) such that 
$J_{nn}\sim a\varphi(n)$, $J_{n\;n+1}\sim b\varphi(n)/2$ as $n\to\infty$, 
where $a$ and $b$ are not simultaneously zero. The function 
$\varphi(x):{\bf R}^+\to {\bf R}^+$
is nondecreasing and such that 
$\lim_{n\to\infty}\varphi(n+x)/\varphi(n)=1$ for any $x\in {\bf R}$; also
let $\lim_{n\to\infty}\varphi(nt)/\varphi(n)=\psi(t)$ exist for $t\in[0,1]$,
and the function $g(t)=dt/d\psi$ be continuous.
We will call the class of such matrices $A_{\varphi(n)}$. 

Let $J(n)$ be the $(n+1)\times(n+1)$ truncated $J$: 
$J(n)=||J_{ik}||_{i,k=0}^n$.
It follows from [\ref{Nev},\ref{Ull}] that for $k=0,1,\dots$\footnote{
The values $\mu_k$ for $b=0$ are obtained from (\ref{mu}) using continuity in
$b$.} 
\begin{equation}
\mu_k\equiv\lim_{n\to\infty}{1\over n}{\rm Tr}
\left[\frac{J(n)}{\varphi(n)}\right]^k=
\frac{1}{\pi}\int_0^1\psi^k(t)dt
\int_{a-|b|}^{a+|b|}\frac{x^k dx}{\sqrt{b^2-(x-a)^2}},
\label{mu}
\end{equation}
Suppose there exists a solution $\rho(z)$ to the moment problem
$\mu_k=\int_{-\infty}^{\infty}z^k \rho(z)dz$, $k=0,1,\dots$.
Since $a$ and $b$ are finite, the function 
$\rho(z)\equiv\rho(z,a,b)$ has a bounded
support which implies that the moment problem is determinate, and hence, 
$\mu_k$, $k=0,1,\dots$ uniquely define $\rho(z,a,b)$. The function 
$\rho(z,a,b)$
is the density of eigenvalues of $J(n)/\varphi(n)$ in the limit $n\to\infty$.
So up to a uniform scaling of the spectrum, $\rho(z,a,b)$ depends only 
on $a/b$.

On the other hand, recall that the orthogonal polynomials satisfy the
recurrence relation [\ref{Chi}]:
\begin{equation} 
xp_n(x)=\alpha_np_{n+1}+\beta_np_n+\gamma_np_{n-1},
\qquad n=0,1,\dots,\qquad \gamma_0p_{-1}=0.
\end{equation}
After the transformation $q_0=p_0$, $q_n(x)=\sqrt{
\frac{\alpha_0\alpha_1\cdots\alpha_{n-1}}{\gamma_1\gamma_1\cdots\gamma_n}}
p_n(x)$, $n=1,2,\dots$, it takes the form
\begin{equation} 
\eqalign{
xq_n(x)=J_{n\;n+1}q_{n+1}+J_{nn}q_n+J_{n\;n-1}q_{n-1},
\qquad n=0,1,\dots\\
J_{nn}=\beta_n,\quad J_{n\;n+1}=J_{n+1\;n}=\sqrt{\alpha_n\gamma_{n+1}},
\quad J_{0\;-1}q_{-1}=0,}
\end{equation}
where $J=||J_{ik}||_{i,k=0}^\infty$ is a symmetric Jacobi matrix. 
As is well known (e.g., [\ref{Chi}]), the zeros of $p_{N+1}(x)$ are 
eigenvalues of $J(N)=||J_{ik}||_{i,k=0}^N$. Thus the asymptotics of the 
zeros of $p_n(x)$ gives the asymptotics of the eigenvalues of $J(n)$. Notice,
that if $J\in A_{\varphi(n)}$ and  $\rho(z,a,b)$ is defined as above, we get
$\rho(z,a,b)=\rho(z,a,-b)$
by the substitution $\tilde{q}_n(x)=(-1)^nq_n(x)$.

Now let us consider the symmetric Jacobi matrices associated with 
the Meixner and
Meixner-Pollaczek polynomials. For Meixner-Pollaczek [\ref{Koe}] we get
\begin{equation}
J_{n\;n+1}=\frac{\sqrt{(n+1)(2\lambda+n)}}{2\sin\phi}\sim{n\over 2\sin\phi};
\qquad
J_{nn}=-(n+\lambda)\cot\phi\sim-n\cot\phi.
\end{equation}
Thus, $J\in A_n$
and $a/b=-\cos\phi$, $\phi\in(0,\pi)$. So $-1< a/b < 1$.

For the Meixner polynomials ($\phi=\ln\sqrt{c}$) 
\begin{equation}
J_{n\;n+1}=\frac{\sqrt{(n+1)(n+\beta)c}}{c-1}\sim{n\over 2\sinh\phi};
\qquad
J_{nn}=\frac{n+(n+\beta)c}{1-c}\sim-n\coth\phi.
\end{equation}
Thus,  $J\in A_n$ and
$a/b=-\cosh\phi$, $\phi\in(-\infty,0)$. So $-\infty<a/b< -1$.

Now it is easy to verify that, up to a sign, an arbitrary matrix $J$ from
$A_n$ for which $|a|\ne|b|$ has the form $rJ_0$, where $r=\sqrt{|a^2-b^2|}$ 
and $J_0$ is a matrix associated with either Meixner-Pollaczek or Meixner 
polynomials defined up to a sign. Obviously,
$\rho(z,a,b)=\rho_0(z/r,a/r,b/r)/r$, where $\rho$ and $\rho_0$ are the
eigenvalue densities of $J(n)/n$ and 
$J_0(n)/n$ as $n\to\infty$, respectively. 

Now we can reformulate Theorem 1 and the first part of Theorem 2.

\noindent{\bf Theorem 3.} {\it Let $J\in A_n$, 
$J_{nn}\sim an$, $J_{n\;n+1}\sim bn/2$ as $n\to\infty$. 
Then the eigenvalue density of $J(n)/n$ in the limit $n\to\infty$
is the following: 

1) $0\le a/b \le 1$, 
\[
\rho(z,a,b)=\cases{
\frac 1{\pi r} {\rm arccosh}\left|{r^2\over zb}+{a\over b}\right|, 
& $z\in[a-b,a+b]$\cr
0, & $z\notin[a-b,a+b]$};
\]

2) $a/b>1$,
\[
\rho(z,a,b)=\cases{1/r, & $z\in[0, a-b]$\cr
\frac 1{\pi r} {\rm arccos}\left(-{r^2\over zb}+{a\over b}\right), 
& $z\in[a-b, a+b]$\cr
0, & $z\notin[0,a+b]$},
\]
where $0\le\arccos(x)\le\pi$, $r=\sqrt{|a^2-b^2|}$.}

In the case $a/b>1$,
more precise information about the zeros in the region where $\rho(z)=1/r$ is
given by the function $\chi(z)$ (Theorem 2).

It was sufficent to consider only the case $a\ge 0$, $b\ge 0$
(substitution $b\to-b$ does not affect the spectrum, and $a\to-a$ inverts it
with respect to zero). In what follows, we always assume $a$ and $b$
nonnegative.
 
For $a=0$ our result coincides with a particular case of the
Nevai-Ullman distribution.
This distribution is the following [\ref{Ull}]:
\begin{equation}
\rho(z,0,b)=\frac{1}{\pi}\int_{|z/b|}^1\frac{g(\omega)d\omega}
{\sqrt{b^2\omega^2-z^2}},\qquad z\in[-b,b].\label{NU}
\end{equation}
For $\varphi(n)=n^\gamma$, $\gamma> 0$, we have
$g(\omega)=\omega^{-1+1/\gamma}/\gamma$.\footnote{
The results for $\gamma=0$ are obtained using continuity.}

It is not difficult to generalize (\ref{NU}) to arbitrary $a$
following the approach of  [\ref{Nev},\ref{Ull}].
It is, however, interesting because of qualitatively new features which appear
when $a>b$. Note that this generalization appears for $\varphi(n)=n^\gamma$
in [\ref{VA1}], where a different approach to its derivation is used.

\noindent{\bf Theorem 4.} {\it Let $J\in A_{\varphi(n)}$, 
$J_{nn}\sim a\varphi(n)$, $J_{n\;n+1}\sim b\varphi(n)/2$ as 
$n\to\infty$. Then the eigenvalue density of $J(n)/\varphi(n)$ in the 
limit $n\to\infty$ is the following: 

1) $0\le a/b \le 1$, 

\noindent
$\rho(z)=
\frac{1}{\pi}\int_{z/(a+b\,{\rm sign}\,z)}^1\frac{g(\omega)d\omega}
{\sqrt{b^2\omega^2 -(z-a\omega)^2}}$  if  $z\in[a-b,a+b]$,
and $\rho(z)=0$ otherwise;

2) $a/b>1$,

\[
\rho(z)=\cases{
\frac{1}{\pi}\int_{(a-b)/(a+b)}^1\frac{g(\omega\frac{z}{a-b})d\omega}
{\sqrt{b^2\omega^2-(a-b-a\omega)^2}}, 
& $z\in[0, a-b]$\cr
\frac{1}{\pi}\int_{z/(a+b)}^1\frac{g(\omega)d\omega}
{\sqrt{b^2\omega^2 -(z-a\omega)^2}}, & $z\in[a-b, a+b]$\cr
0, & $z\notin[0,a+b]$}.
\]
}

Note that for $\varphi(n)=n^\gamma$, $\gamma>0$ we can write case 2
in a simpler form:
\[
\rho(z)=\cases{h(a-b)\left(\frac{z}{a-b}\right)^{-1+1/\gamma}, 
& $z\in[0, a-b]$\cr
h(z),& $z\in[a-b, a+b]$\cr
0, & $z\notin[0,a+b]$},
\]
where
\[
h(z)=
\frac{1}{\pi\gamma}\int_{z/(a+b)}^1\frac{\omega^{-1+1/\gamma}d\omega}
{\sqrt{b^2\omega^2 -(z-a\omega)^2}}.
\]

{\it Proof} is analogous to that given by Ullman in the case $a=0$.
Rewrite (\ref{mu}) in the form:
\begin{equation}
\mu_k={1\over\pi}\int_0^1 g(\psi) d\psi
\int_{a-b}^{a+b}\frac{(\psi x)^k dx}{\sqrt{b^2-(x-a)^2}}.\label{mu2}
\end{equation}
Consider case 1. Changing the variables $z=\psi x$, $\omega=\psi$ in the
double integral, we have:
\begin{equation}
\mu_k={1\over\pi}\int_{a-b}^{a+b}z^k dz
\int_{z/(a+b\,{\rm sign}\,z)}^1\frac{g(\omega)d\omega}
{\sqrt{b^2\omega^2 -(z-a\omega)^2}},\label{c1}
\end{equation}
which proves case 1. Similarly, we prove case 2.


Theorem 1 and the first part of Theorem 2 (that is Theorem 3) are 
corollaries of Theorem 4 when $\varphi(n)=n$. 

Finally, note an important and, in its generality, very difficult problem:
what weights of orthogonal polynomials lead to the matrices in 
$A_{\varphi(n)}$?
A recent review of related results is given in [\ref{Lub}]. 
In this sense, the case $|a|>|b|$ is studied less than the case $|a|<|b|$.

\section{Acknowledgements}
I am grateful to Renato \'Alvarez Nodarse, Mourad Ismail, Alphonse Magnus, 
Paul Nevai, and Walter Van Assche for useful correspondence.
I am indebted to O. B. Zaslavsky for an important discussion at the early 
stage of this project.

\end{document}